\documentclass[aps,pra,twocolumn,groupedaddress,showpacs,floatfix]{revtex4}

\usepackage{graphicx}

\usepackage{epstopdf}

\usepackage{amsmath,amssymb}

\def\openone{\leavevmode\hbox{\small1\kern-3.3pt\normalsize1}}

\newcommand{\Op}[1]{\boldsymbol{\mathsf{\hat{#1}}}}

\newcommand{\Fkt}[1]{\,\mathsf {#1}}

\ifx\Tr\renewcommand{\Tr}{\Fkt{Tr}}
\else\newcommand{\Tr}{\Fkt{Tr}}
\fi

\begin{document}

\title{Photoassociation and coherent transient dynamics in the
  interaction of ultracold rubidium atoms with shaped femtosecond
  pulses - II. Theory}  

\author{Andrea Merli}

\author{Frauke Eimer}

\author{Fabian Weise}

\author{Albrecht Lindinger}

\affiliation{Institut f\"ur Experimentalphysik, Freie Universit\"at
  Berlin, Arnimallee 14, D-14195 Berlin, Germany} 

\author{Wenzel Salzmann}

\author{Terry Mullins}

\author{Simone G\"otz}

\author{Roland Wester}

\author{Matthias Weidem\"uller}
\altaffiliation[now at: ]{Physikalisches Institut, Ruprecht-Karls-Universit\"at
  Heidelberg, Philosophenweg 12, D-69120 Heidelberg, Germany}

\affiliation{Physikalisches Institut, Universit\"at Freiburg, Hermann
  Herder Str. 3, D-79104 Freiburg i. Br.} 

\author{Ruzin A\v{g}ano\v{g}lu}

\author{Christiane P. Koch}

\affiliation{Institut f\"ur Theoretische Physik, Freie Universit\"at
  Berlin, Arnimallee 14, D-14195 Berlin, Germany} 

\email[corresponding author: ]{ckoch@physik.fu-berlin.de}

\begin{abstract}
Photoassociation of ultracold rubidium atoms with femtosecond laser
pulses is studied theoretically. The spectrum of the pulses is cut off
in order to suppress pulse amplitude at and close to the atomic resonance
frequency. This leads to long tails of the laser
pulse as a function of time giving rise to coherent transients in
the photoassociation dynamics. They are studied
as a function of cutoff position and chirp of the pulse. Molecule
formation in the electronically excited state is attributed to
off-resonant excitation in the strong-field regime.
\end{abstract}


\maketitle

\section{Introduction}

Cooling, trapping and manipulation of atoms and molecules in the
ultracold regime ($T \le 100\,\,\mu$K) represents one of the most
active research fields in contemporary atomic, molecular, and optical
physics. 
Photoassociation is the formation of molecules when two colliding
atoms are excited by laser light into bound vibrational levels
of an electronically excited state. Using continuous-wave lasers,
photoassociation of ultracold atoms 
can be employed to study collisional properties of ultracold atoms, in
particular long-range potential energy curves \cite{JonesRMP06}.
If followed by spontaneous emission, it also serves to form ultracold molecules in
their electronic ground state \cite{FrancoiseReview}. 

The use of short
laser pulses for photoassociation of ultracold atoms has been suggested in
theory work for about a decade
\cite{VardiJCP97,vala2001,ElianePRA04,KochPRA06a,KochPRA06b,KochJPhysB06,ShapiroPRA07,UliMyJPhysB06}. 
Initially, the main goal was to find a photoassociation scheme that
would allow for larger formation rates of excited state molecules
\cite{vala2001,ElianePRA04}. In order to coherently produce ultracold
molecules in their electronic ground state, the pump-dump scenario was
adapted to photoassociation \cite{KochPRA06a,KochPRA06b,KochJPhysB06}:
A pump pulse creates a vibrational wavepacket on the electronically
excited state which travels toward its inner turning point where it is
sent to the ground state by the dump pulse. 
Other coherent control schemes such as adiabatic
passage \cite{ShapiroPRA07} and pulse optimization using
genetic algorithms were investigated as well \cite{UliMyJPhysB06}.
Coherent photoassociation using short laser pulses
is closely linked to the ultimate goal of ultracold molecule
formation, the
 stabilization of the molecules into their
vibrational ground 
state \cite{DanzlSci08,NiSci08,DeiglmayrPRL08,LangPRL08}.
This has been discussed in theory work in the context of
optimal control 
\cite{MyPRA04} and coherent pulse accumulation \cite{PeerPRL07}.

In the theoretical studies on short-pulse photoassociation,
pulses with a spectral bandwidth of a few wavenumbers 
were suggested. This choice is
motivated by the requirement of exciting a
narrow band of transitions with large free-bound Franck-Condon factors
close to the atomic resonance without exciting atomic transitions. 
This spectral bandwidth corresponds to transform-limited pulse
durations of a few picoseconds. However, 
pulse shaping capabilities for such pulses have yet to be developed.

Femtosecond
pulses on the other hand can be shaped with a number of techniques, 
but they have a very broad spectrum, addressing both
molecular and atomic transitions. Since the probability of atomic
transitions is several orders of magnitude larger than that of
photoassociation transitions, the pulse spectral amplitude at the atomic
resonance frequency needs to be
completely suppressed in order to record any molecular signal. This is
most easily achieved by putting a knife edge into the Fourier plane of
the pulse shaper \cite{SalzmannPRA06}. The corresponding sharp cutoff in
the spectrum of the femtosecond pulse leads to long tails of the
pulse as a function of time and induces coherent transient dynamics
\cite{NiritPRL05,GirardPRL06}. 
The transients are reflected in
oscillations of the molecular population and can be measured by
ionization detection of the excited state molecules \cite{SalzmannPRL08}. 
Photoassociation of ultracold atoms with shaped femtosecond
laser pulses has thus been demonstrated 
recently in a break-through experiment \cite{SalzmannPRL08}.

\begin{figure}[tb]
\begin{center}
\includegraphics[width=\columnwidth]{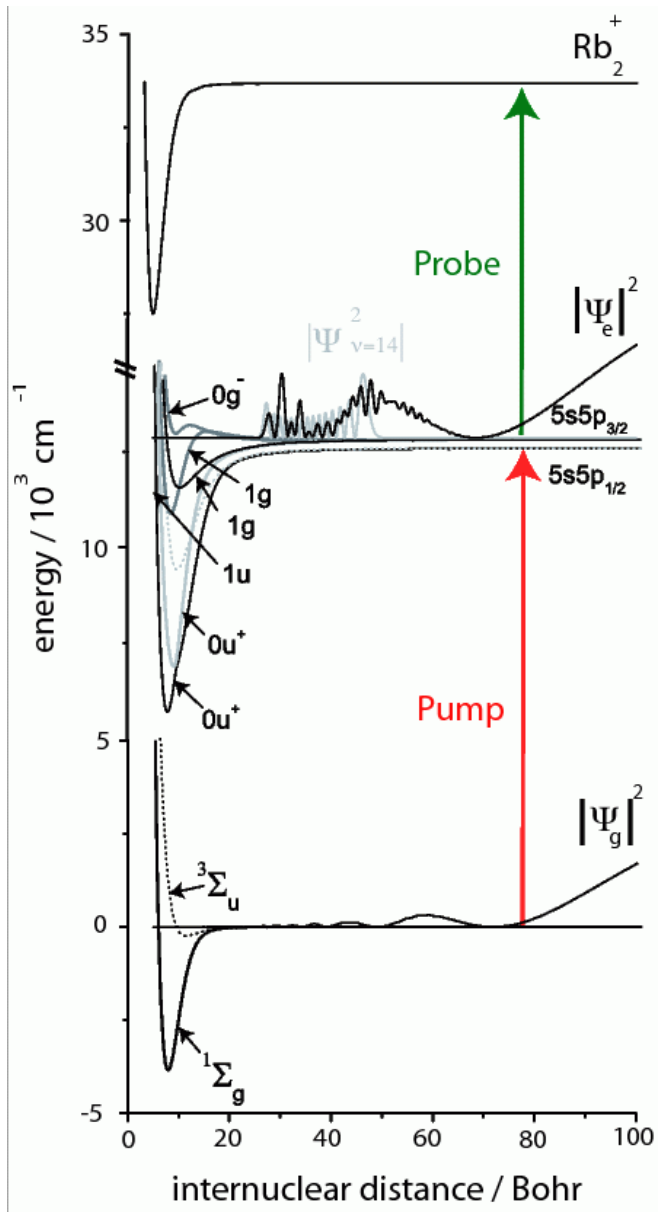}
\caption{(color online) %
  Scheme of the femtosecond pump-probe sequence involving the Rb$_{2}$
  potential energy curves of the electronic ground 
  state, the attractive potentials of the first electronically excited
  state correlating to the 5s+5p$_{1/2}$ and
  5s+5p$_{3/2}$ asymptotes, and the cationic state of Rb$_{2}^+$.
  The initial scattering state, the wavefunction amplitude after the
  pump pulse  and one sample excited state vibrational 
  wavefunction ($v=14$ of the $0_g^-(5s+5p_{3/2})$ state)  
  are also depicted.
  }
\label{fig:ppschema}
\end{center}
\end{figure}

This paper presents a detailed explanation of the experimental results
obtained by femtosecond photoassociation of ultracold rubidium atoms
with shaped pulses based on theoretical
calculations.  The details of the experimental setup as well as the
major part of the experimental results is found in
Ref.~\cite{part1}. 
Figure \ref{fig:ppschema} depicts the proposed photoassociation mechanism
including the potential energy curves considered in the
calculations. A first laser field (the pump pulse, indicated by the red arrow)
excites two atoms from their  
ground state (5s+5s) into long-range bound molecular levels of an
electronically excited state (correlated to the 5s+5p$_{1/2}$ or
5s+5p$_{3/2}$  asymptotes).
The  initial scattering state with a relative
scattering energy corresponding to the trap temperature of
100$\,\mu$K is shown on the ground state. The excited state
wavefunction after the pulse is over as well as one
exemplary excited state vibrational wavefunction are displayed in the
first excited state of Fig.~\ref{fig:ppschema}.
In order to detect the excited state molecules,
a second laser field
(the probe pulse, indicated by the green arrow)
transfers them to the ionic state. 

The paper is organized as follows: Section~\ref{sec:TA} briefly
reviews the theoretical description of two rubidium atoms excited by a
photoassociation laser pulse. The photoassociation mechanism based on
strong-field off-resonant excitation is described in detail in
Sec.~\ref{sec:PA} while the coherent transient dynamics are studied in 
Sec.~\ref{sec:cohtrans}. Theoretical and experimental results are
compared in Sec.~\ref{sec:comparison} and Sec.~\ref{sec:concl}
concludes. 

\section{Theoretical Approach}
\label{sec:TA}

In order to gain detailed insight into the mechanism of femtosecond
photoassociation, the pump excitation step was simulated in quantum
dynamical calculations of the light-molecule  
interaction. The pump pulse transfers population between the ground
and first excited electronic states. The probe step was not explicitly
taken into account. Rather, it was assumed that any population of the
first excited state is ionized alike. 

In a two-channel picture, the Hamiltonian describing two rubidium
atoms subject to excitation by a laser field in the dipole and
rotating-wave approximations is given by
\begin{equation}\label{eq:H}
  \Op{H}=  \begin{pmatrix}
    \Op{T}+V_{g}(\Op{R})& \Op{\mu} E(t) \\ 
    -\Op{\mu} E(t)^{*} &\Op{T}+V_{e}(\Op{R})-\Delta
  \end{pmatrix}.
\end{equation}
Here, $\Op{R}$ denotes the internuclear distance, $\Op{T}$ the kinetic
energy operator and $V_{g/e}$ the ground and excited state 
 potential energy curves. 
The parameter $\Delta$=$\hbar$($\omega_{0}-\omega_{L}$) represents the
detuning of the pump pulse central frequency relative to the
$5s+5p_{1/2}$ and $5s+5p_{3/2}$ asymptotes, respectively.  
The off-diagonal elements of the Hamiltonian describe the electric
dipole coupling with $\Op{\mu}$ denoting the transition dipole and
$E(t)$ the electric field of the pump pulse. The $R$-dependence of the
transition dipole was neglected which is justified for excitation at
long range.  

A two-channel picture
assumes that no couplings between electronic states, for example due
to spin-orbit interaction, play a role on the timescale of the
dynamics. It is then possible to consider pairwise combinations 
of ground and excited electronic states in Eq.~(\ref{eq:H}).
An analysis of the dynamics has confirmed that this assumption is 
justified: The dynamics are purely long-range and dominated
by electronic transitions due to the laser field.

The potential energy curves are
taken from Ref. \cite{park2001} and matched to the long-range
dispersion potentials, $(C_3/R^3+)C_6/R^6+C_8/R^8+C_{10}/R^{10}$ with the $C_n$
coefficients found in Refs. \cite{MartePRL02,GuterresPRA02}.
Potential energy curves in the Hund's case (c) representation were
obtained by approximating the spin-orbit couplings by their asymptotic
values and subsequent diagonalization, see e.g. Ref.~\cite{WangJCP96}. 
Of all potential curves correlating to the $5s+5p_{1/2}$ and
$5s+5p_{3/2}$ asymptotes, those with attractive character and
dipole-allowed transitions were taken into account ($0_g^-$, $1_g$ and
$0_u^+$ correlating to the 5s+5p$_{1/2}$ asymptote and $0_g^-$, $1_g$,
$0_u^+$ and $1_u$ correlating to the  5s+5p$_{3/2}$ asymptote).
Moreover simulations were also performed for two repulsive potential
energy curves, those of the $1_u(5s+5p_{3/2})$ and $2_g(5s+5p_{3/2})$
states. 
Note that the excited state potentials differ by their
effective $C_3$ coefficients and strengths of transition dipole
moment. 

The Hamiltonian, Eq.~(\ref{eq:H}), is represented on a mapped Fourier grid
\cite{SlavaJCP99,WillnerJCP04}. Vibrational wavefunctions
such as the one for $v=14$ of the $0_g^-(5s+5p_{3/2})$ excited state
shown in Fig.~\ref{fig:ppschema} are obtained
by diagonalizing  the Hamiltonian with $E(t)$ set to zero. The
photoassociation dynamics were studied by numerically solving the
time-dependent Schr\"{o}dinger equation (TDSE),
\begin{equation}
  \label{eq:TDSE}
  i\hbar \frac{\partial}{\partial t}
  \begin{pmatrix}
    \Psi_g(R;t) \\  \Psi_e(R;t)
  \end{pmatrix} = \Op{H}
    \begin{pmatrix}
    \Psi_g(R;t) \\  \Psi_e(R;t)
  \end{pmatrix} \,,
\end{equation}
using the Chebyshev propagator \cite{kosloff1988}. $\Psi_{g/e}(R;t)$
denotes the ground (excited) state component of the time-dependent
wavefunction. The initial state is taken to be the scattering state
with scattering energy closest to $E_{scatt}/k_B =T_{trap}$ where
$k_B$ denotes the Boltzmann constant and  $T_{trap}$ the temperature of
the trap. In principle an average over all thermally populated
scattering states needs to be performed \cite{KochJPhysB06}. However,
since the dynamics are dominated by electronic transitions, the
different nodal structures of the scattering states at very
large distances do not play a role and simulations for a single
initial state are sufficient. 

\begin{figure*}[tb]
\begin{center}
\includegraphics[width=17cm]{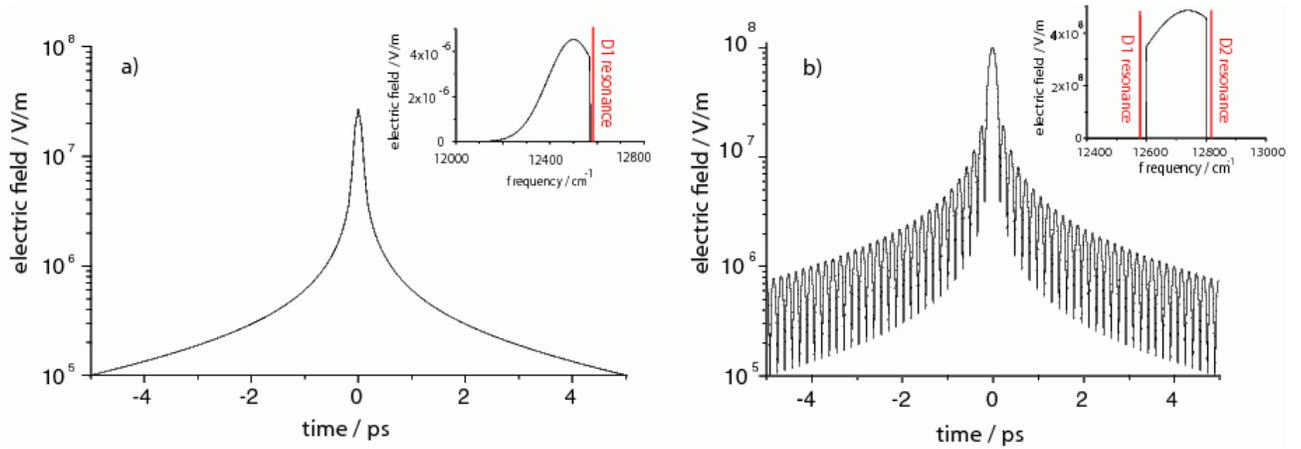}
\caption{(color online)
  Temporal field envelope of cut pump pulse spectrum (shown in the
  insets)  below the
  D1 (a) and D2 (b) atomic resonance. Tails to both
  sides of the peak are caused by the spectral cutoff.}
\label{fig:pulsform}
\end{center}
\end{figure*}

The pulses entering Eqs.~(\ref{eq:H}) and (\ref{eq:TDSE}) were
obtained similar to the experimental procedure. 
In the experiment an optical low pass filter was employed in the
Fourier plane of the 
zero dispersion compressor to manipulate the spectral shape of the
pump pulse close to the D1 or D2 atomic line in such a way, that 
all frequencies above the chosen atomic resonance, the respective
resonance frequency itself, and some frequencies 
below this particular line were cut off. This is done in order to
prevent the loss of the captured atoms due 
to resonant three-photon ionization and photon pressure
\cite{SalzmannPRA06}. In the case of excitation near the D2 line, 
perturbing frequencies near the D1 atomic resonance due to the 
large spectral bandwith of the femtosecond pulse 
were also removed from the pump pulse spectrum.
In order to model those experimental pump pulses, a Gaussian
transform-limited laser field is subjected to one or two cuts of spectral
amplitude in the frequency domain. The sharp cuts
lead to very long transients of the pulse in the time-domain
representation, cf. Fig.~\ref{fig:pulsform}, i.e. the temporal behavior
strongly deviates from a transform limited profile. The pulse has long tails, 
reaching out to a few picoseconds before and after the maximum as
shown in Fig.~\ref{fig:pulsform}.
If two cuts are applied, the tails show a beat pattern that depends on
the specific cut positions, cf. Fig.~\ref{fig:pulsform}(b).
Within the tails, the 
electric field oscillates with a frequency that matches the spectral
cutoff frequency, cf. Fig.~\ref{fig10_dipole_interaction}.
The influence of the specific shape of the pump
pulses in time and frequency domain on the dynamics
will be discussed in Sec.~\ref{sec:cohtrans}.
The parameters defining the pulse are 
the central wavelength, spectral width, 
energy, linear chirp rate and spectral cutoff positions.

The time-dependent population of the first electronically excited
state is calculated as weighted sum of the possible transitions
from the singlet ground and lowest triplet state with the ratio 1:3.
In order to take the finite duration of the
probe pulse into account, the measured ion 
signal was obtained by convoluting the time-dependent excited state
population with a Gaussian of 500$\,$fs full-width at half maximum
(FWHM)\footnote{All full widths at half maximum are taken with respect
  to the intensity profile.}.  

\section{Photoassociation}
\label{sec:PA}

\begin{figure*}[tb]
\includegraphics[width=17cm]{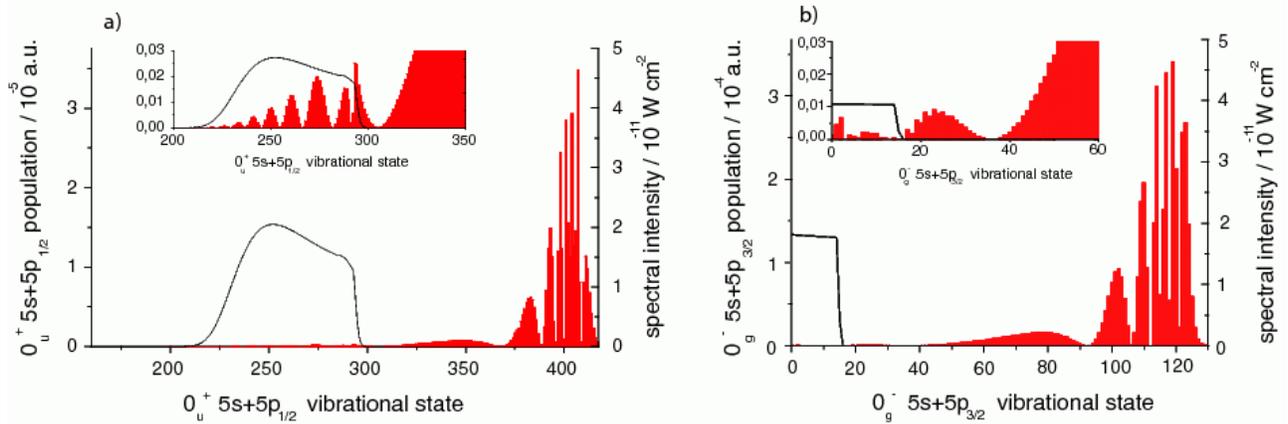}
\caption{(color online)
  Calculated population distribution in the
  excited $0_u^+(5s+5p_{1/2})$ (a) and $0_g^-(5s+5p_{3/2})$ (b)
  states.  
  The solid line displays the spectral pump pulse amplitude where the
  transition energies from the initial ground state scattering state
  to the bound excited state vibrational levels were matched to the
  pump pulse spectrum.}
\label{fig:vibstate} 
\end{figure*}

The photoassociation dynamics is illustrated in Fig.~\ref{fig:ppschema}
by the nuclear wavefunctions, $\Psi_g(R;t)$ and $\Psi_e(R;t)$,
in the ground and excited state potentials, when the pulse is over, 
at about $t=18\,$ps  after the pump pulse
maximum. The $a\Sigma_u^+(5s+5s)$ lowest triplet state and
$0_g^-(5s+5p_{3/2})$ excited state were employed in this example. 
Pump pulse parameters 
like pulse energy and cutoff position correspond to those used in
the actual experiment. Specifically, the cutoff positions for the example
displayed in Fig.~\ref{fig:ppschema} and Fig.~\ref{fig:vibstate}(b),
were $-15\,$cm$^{-1}$ and $-218.3\,$cm$^{-1}$ below
the D2 resonance, the central wavelength $\lambda_L=785\,$nm,
spectral bandwidth $\Delta\lambda_L=25\,$nm, laser power $P=30\,$mW,
repetition rate  $100\,$kHz
and beam radius $r_B=300\,\mu$m, corresponding to a peak field
amplitude of 99$\,$MV/m. 
Inspection of Fig.~\ref{fig:ppschema} reveals that no nuclear dynamics
has taken place on the time scale of the pulse including its long
tails: The excited state wave packet,
created by the pump pulse, completely
reflects the nodal structure of the initial scattering state at long
range. 
This is not surprising since the vibrational periods of weakly bound
excited state levels are on the order of several tens of picoseconds,
while the ground state dynamics evolves even more slowly.

The excited state wave packet at a certain time can be projected onto
the vibrational eigenfunctions of the
corresponding excited state potential. This yields
the vibrational distribution and proves that photoassociation,
i.e. the excitation of free colliding atoms into bound levels of an
excited state, has taken place. The vibrational distribution after the
pulse is shown for the $0_u^+(5s+5p_{1/2})$
state in Fig.~\ref{fig:vibstate}(a) and for the $0_g^-(5s+5p_{3/2})$
state in  Fig.~\ref{fig:vibstate}(b). The cutoff position was chosen
to be $15\,$cm$^{-1}$ below the respective atomic resonance,
i.e. below the D1 line frequency in Fig.~\ref{fig:vibstate}(a) and
below the D2 line frequency in  Fig.~\ref{fig:vibstate}(b).
In Fig.~\ref{fig:vibstate}(a), 
the vibrational distribution is characterized by a maximum near
$v'=400$. These vibrational levels have large bond lengths
on the order of $1000\,a_0$ and low binding energies of
about 1$\,$GHz or 0.03$\,$cm$^{-1}$. They lie energetically
above the spectral cutoff of the pulse.
Their population therefore necessarily arises from off-resonant
excitation. This is caused by the high peak intensity of the pump pulse.
Only a small fraction of the total excited state population resides
in vibrational levels that are resonant within the pump pulse spectrum. For
a spectral cutoff detuned by -15\,cm$^{-1}$ from the atomic line,
resonant excitation occurs for vibrational levels
below $v'=290$, cf. inset of Fig.~\ref{fig:vibstate}(a).
Compared to the weakly bound, long range levels around $v'=400$,
their population is suppressed by two orders of magnitude. This is 
due to much smaller free-bound Franck-Condon factors.
The distribution among these levels reproduces the weak-field result,
i.e. it is proportional to the Franck-Condon factors multiplied by 
the spectral envelope of the pump pulse \cite{UliMyJPhysB06,shapirobook}.
The apparant contradiction of producing within one calculation
weak-field (below $v'=290$) and strong-field (near $v'=400$) results is
easily resolved by noting that the decisive quantity is the Rabi
frequency, i.e. the product of transition dipole and 
electric field. Since the Franck-Condon factors vary over several
orders of magnitude, qualitatively different results are obtained for
different regions of the excited state vibrational spectrum. 

A pattern similar to that of the D1 line results 
is observed for the D2 line, cf. Fig.~\ref{fig:vibstate}(a) and (b). In
this case, the excited state potential, the purely long range
$0_g^-(5s+5p_{3/2})$ state, is only about 28$\,$cm$^{-1}$ deep and 
resonant excitation occurs for vibrational levels
below $v'=15$. Due to the shallowness of the potential, only a small
part of the pulse spectrum serves for excitation, cf. the inset of
Fig.~\ref{fig:pulsform}(b). If an excited state
potential with a deep minimum such as that of the $1_g(5s+5p_{3/2}$
state is employed (data not shown), a picture equivalent to
Fig.~\ref{fig:vibstate}(a) emerges.  In both cases the vibrational
spectrum is dominated by 
levels which are off-resonantly excited, above $v'=90$ for the
$0_g^-(5s+5p_{3/2})$ state (with binding
energies of less than 0.05$\,$cm$^{-1}$ or 1.5$\,$GHz). 
It is instructive to compare the amplitudes of the excited state
vibrational level with $v'=14$ and the 
excited state wave packet after the pulse, shown in 
Fig.~\ref{fig:ppschema}, to the vibrational distributions of
Fig.~\ref{fig:vibstate}(b).  
The short range part of the wave packet at internuclear distances
$R\le 50\,a_0$ corresponds to vibrational levels with $v'\le 14$. The
total excited state population is dominated
by the weakly bound levels and electronically excited
continuum states.  Population of the latter,
however, corresponds to excitation of the initial colliding atom
pair to an excited pair of free atoms and  cannot be regarded as
photoassociation.

\section{Coherent transient dynamics}
\label{sec:cohtrans}

\begin{figure}[tb]
\begin{center}
\includegraphics[width=\columnwidth]{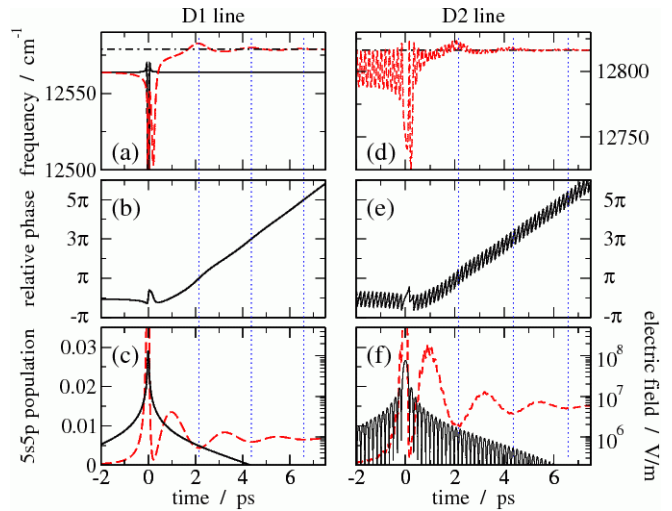}
\caption {
  (color online) 
  Details of ultrafast dipole dynamics. a+d:
  instantaneous frequency of the pump pulse (solid black line), frequency of
  induced dipole (dashed red line), position of the D1, resp. D2,
  resonance frequency (dash-dotted line). b+e: relative phase between
  dipole and light field. 
  c+f: electric field envelope of pump pulse (solid black line), 
  excited state population (dashed red line) for the $0_u^+(5s+5p_{1/2})$
  state (c) and $0_g^-(5s+5p_{3/2})$ (f). The vertical dotted blue
  lines indicate the position where the relative phase crosses
  $(n+1)\pi$, $n=0,1,2$.} 
\label{fig10_dipole_interaction}
\end{center}
\end{figure}

Solving the time-dependent Schr\"odinger equation allows for a
detailed study and analysis
of the dynamics induced by the pump
pulse. Figure~\ref{fig10_dipole_interaction} displays the 
excited state population, $\int dR |\Psi_e(R;t)|^2$, and the
frequency of the induced dipole, i.e. the derivative of the
time-dependent phase, $\phi_{\langle\mu\rangle(t)}$,  of $\int dR
\Psi^*_g(R;t)\Op{\mu}\Psi_e(R;t)\rangle$
(red dashed lines in  
Fig.~\ref{fig10_dipole_interaction}~a, c, d, f) as a function of time and
compares them to the envelope and the instantaneous frequency
of the pump pulse  (black solid lines in
Fig.~\ref{fig10_dipole_interaction}~a, c, d, f). 
Also shown is the relative phase between the transition dipole and the
field, i.e. $\phi_{\langle\mu\rangle(t)}-\phi_{E(t)}$ 
(solid black line in Fig.~\ref{fig10_dipole_interaction}~b).
Characteristic oscillations of the excited state population are
observed which are due to the long tails of the electric field with
cut spectrum, cf. Fig.~\ref{fig10_dipole_interaction}~c. They
correspond to the coherent transients \cite{shapirobook} that have
been observed before in the excitation
of atoms with short pulses, see
e.g. Refs.~\cite{NiritPRL05,GirardPRL06}.
Note that in our case, a molecular signal is observed in the
ionization detection \cite{part1}.

The transient interaction of the
weakly bound molecule and the pulse is governed by the induced
transition dipole, in particular by the relative phase between the
induced dipole and the field. The dynamics obtained for a single
spectral cut, i.e. for excitation below the D1 line (left-hand side of
 Fig.~\ref{fig10_dipole_interaction}), is discussed
first. In this case, 
the instantaneous frequency of the laser pulse, shown by the
black solid line in Fig.~\ref{fig10_dipole_interaction}~a,
coincides with the cutoff
frequency before and after the pulse maximum but dips to
12250$\,$cm$^{-1}$ during the main part of the pulse. The depth of the
dip is determined by the spectral bandwidth of the pulse. 
The temporal tails to both sides
of the main pulse peak lead to significant field amplitude that
oscillates with the cutoff frequency, cf. black solid line in
Fig.~\ref{fig10_dipole_interaction}~a. During the main peak of the pulse,
around $t=0$, a dipole is induced by off-resonant excitation. It 
is at first driven by the strong electric field and follows its
oscillation. After the main peak of the pulse is over, the coupling is
reduced and the dipole oscillates with its intrinsic frequency close
to the D1 resonance frequency, cf. red dashed line in
Fig.~\ref{fig10_dipole_interaction}~a. This is due to the excited state
population being dominantly in vibrational levels 
close to the dissociation limit, cf. Section
\ref{sec:PA}.
The oscillation of the excited state population goes through
one period as the relative phase between dipole and field changes
by 2$\,\pi$, see the dotted blue lines and the red dashed curve in
Fig.~\ref{fig10_dipole_interaction} c. The dotted blue lines indicate
the times when the relative phase crosses odd integer multiples of
$\pi$. These times coincide with the minima in the oscillation of the
excited state population. 
In the long pulse tail, this oscillation results from a beating
between the transition dipole and the field that is oscillating with
the cutoff frequency. As the long tail of the pump pulse 
decays (black solid line in Fig.~\ref{fig10_dipole_interaction}~c), 
 the oscillations of the transition dipole subside as
well. This damps effectively the oscillations in the excited state
population since the energy exchange between pulse and molecule
vanishes. 

The dynamics observed for a pulse to which two spectral cuts are
applied, i.e. for excitation below the D2 resonance frequency
(right-hand side of Fig.~\ref{fig10_dipole_interaction}), is overall
similar to that obtained for excitation below the D1 resonance
frequency. However, two spectral cuts lead to a distinct beat pattern
in the amplitude and phase of the electric field which carries over
to the dynamical observables transition dipole and excited state
population. The beat pattern is determined by the positions of the
two cuts relative to the carrier frequency of the field. 
The induced dipole oscillates around the cut position before the main
peak of the pulse. This oscillation is of little relevance, however,
since the amplitude of the induced dipole which is related to the
excited state population is extremely small. It dips down to
12725$\,$cm$^{-1}$ during the main peak of the pulse and oscillates
with its intrinsic frequency close to the D2 resonance superimposed
with the fast beat pattern after the main peak is over. 
While the beat patterns
are easily resolved in the calculations where a time step of 1$\,$fs
was employed (the results shown in
Fig.~\ref{fig10_dipole_interaction} are not convoluted with a Gaussian
representing the effect of the probe pulse), they require a probe pulse
with FWHM of 100$\,$ or less and a sufficiently good signal-to-noise
ratio to be observed experimentally in e.g. the excited state
population. In the experiment discussed here 
the comparatively long duration of the probe pulse does not allow for
the observation of this beat pattern.

The coherent transients are solely due to electronic transitions as
explained in detail below in Section~\ref{sec:comparison}. 
Their dynamics can therefore also be modelled in the
framework of dressed states of a two-level system. This approach
has the advantage of yielding 
analytic expressions which describe the coherent interactions
causing the dipole modulations \cite{albert2008}.

\section{Comparison of theoretical and experimental results} 

\label{sec:comparison}

\begin{figure*}[tb]
\includegraphics[width=17cm]{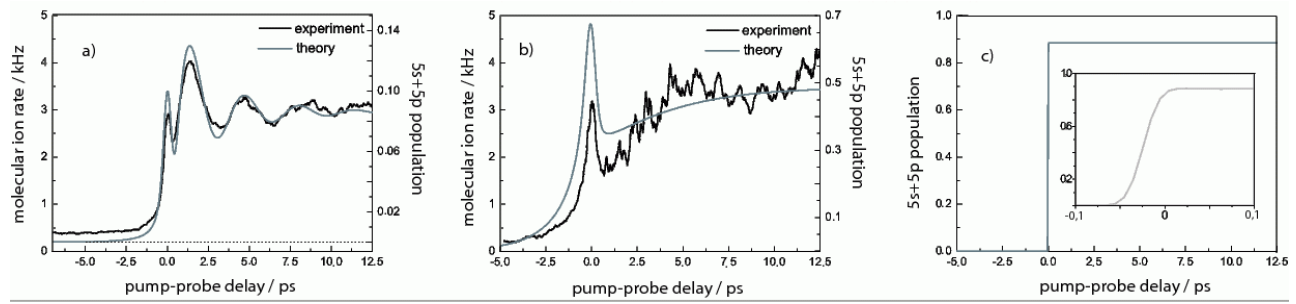}
\caption{\label{fig:vergleichkante}
  Experimental (black) and theoretical (grey)
  pump-probe signals for pump pulses with spectral cutoff (a)
  10$\,$cm$^{-1}$ red to the D2 line resonance,  (b)
  directly on the resonance frequency, and (c) theoretical signal
  for a pump pulse
  without spectral cutoff (transform-limited pulse).
  }
\end{figure*}
A detailed comparison of the quantum dynamical calculations to the
experimental results is presented in the following.
A typical pump-probe signal
recorded in an experiment with excitation frequency near the D2 line is
depicted by the black curves in Fig.~\ref{fig:vergleichkante}, while
the grey curves 
represent the excited state population obtained in the calculations after
convolution with the probe pulse (cf. Sec.~\ref{sec:TA}).  
For a chosen cutoff position below the D2 resonance (here 10$\,$cm$^{-1}$), a
constant molecular ion signal is measured at negative time delays
where the probe pulse precedes the pump pulse,
cf. Fig.~\ref{fig:vergleichkante}a. 
For zero delay, the pump and probe pulses coincide in time
and a peak of about 0.5$\,$ps width is observed. For positive time
delays where the pump pulse precedes the probe 
pulse, an increased molecular ion signal occurs with characteristic
oscillations that are fully damped after a few picoseconds.
An excellent agreement between the measured
pump-probe data and the simulated excited state 
population in the attractive potentials of the $5s+5p_{3/2}$ is
observed for positive delays. The finite offset of the experimental
signal at negative delays is absent in the theoretical curve. 
The calculations are done for a single pump pulse, while the
experiments operate with a high repetition rate, 100$\,$kHz. The
signal at negative delays is therefore attributed to 
molecules in the electronic ground state that were produced by the
previous pump pulse and spontaneous emission \cite{SalzmannPRL08}.  
This effect is not taken into account in the calculations. 

When the atomic resonance is included in the pump pulse spectrum, 
 Fig.~\ref{fig:vergleichkante}b, the 
total pulse energy needs to be reduced by 50\% with an optical gray
filter to avoid a too strong perturbation of the ultracold cloud.
In this case, a higher ion signal compared to that obtained with a cut
to the red of the resonance frequency
is observed for positive time
delays. The characteristic oscillations at positive delay
disappear almost completely, cf. Fig.~\ref{fig:vergleichkante}b. 
When a transform-limited pulse is considered in the calculation, the
population is quickly transferred to the excited state, and no
oscillations are observed in Fig.~\ref{fig:vergleichkante}c (the
inset shows a zoom of 200 fs around zero delay). In the
experiment, a transform-limited pulse led to a destruction of the
sample. 
Fig.~\ref{fig:vergleichkante} illustrates that 
the transient oscillations depend
strongly on the spectral cutoff of the pump pulse spectrum. 

\begin{figure*}[tb]
\includegraphics[width=15cm]{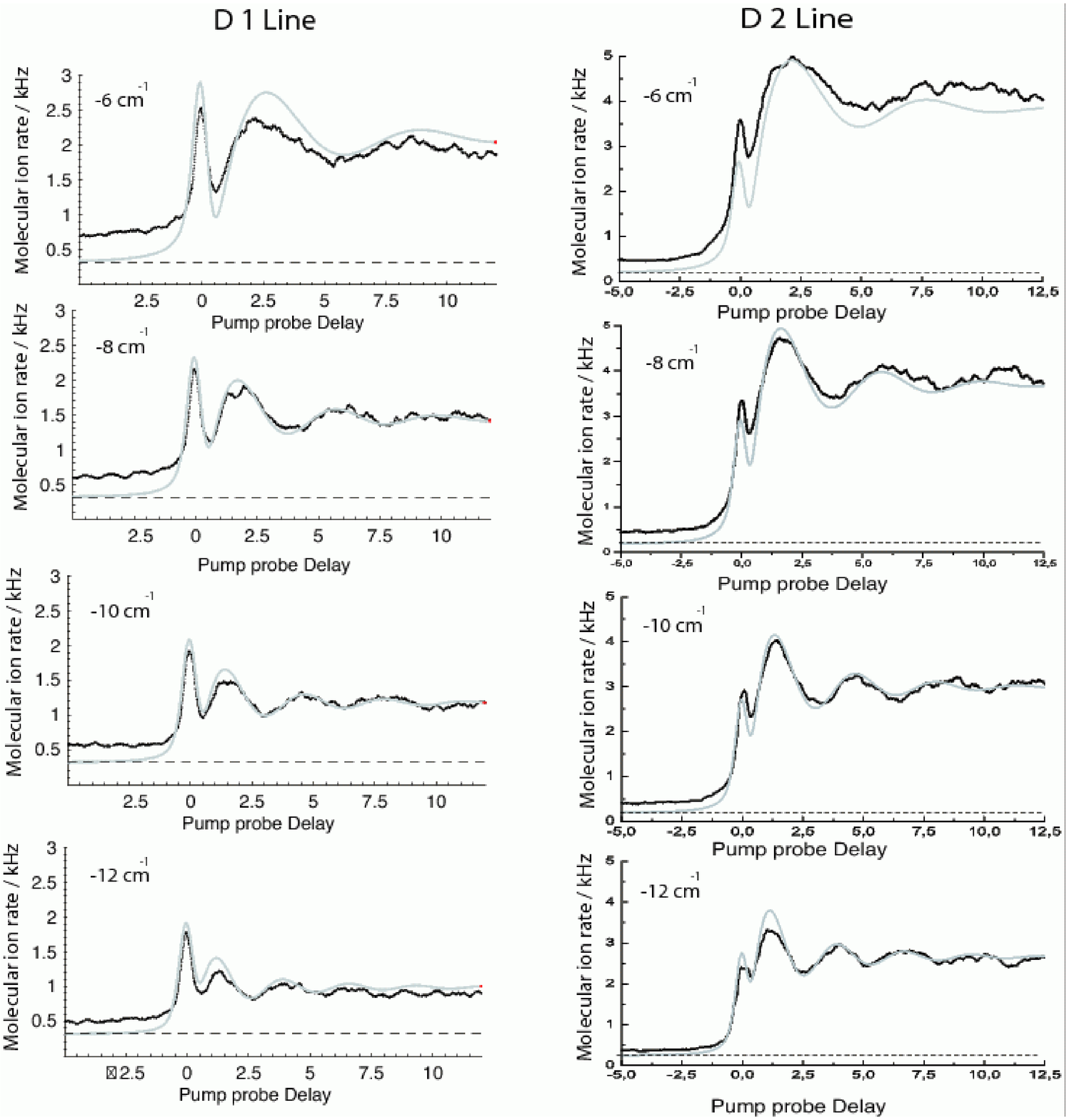}
\caption{\label{fig:Kantenscan1}
  Experimental (black) and theoretical (gray) pump-probe signals 
  for  pump pulses with different 
  spectral cutoff positions red to the D1 line resonance
  ($5s+5p_{1/2}$-asymptote, left)  and D2 line resonance
  ($5s+5p_{3/2}$-asymptote, right), respectively.  
  The constant level of molecules measured by irradiating with the
  probe pulse only is indicated by the dashed lines, and the zero of
  the theoretical results is shifted to coincide with this level.
} 
\end{figure*}
Fig.~\ref{fig:Kantenscan1} 
shows the time-resolved molecular ion signal for different
spectral cutoffs below the D1 and D2 line resonances in comparison
with the corresponding quantum dynamical calculations. 
Generally, the pump-probe spectra obtained by exciting  the
ultracold ensemble near the D1 and D2 line resonances are very
similar.
A notable difference is observed in the  Rb$_{2}^{+}$ count rate which
is larger for excitation near the D2 line and positive delays.
This is attributed to the larger dipole moment for the $5s\rightarrow
5p_{3/2}$ transition. When the sample is irradiated by the probe pulse
only, a constant level of molecules is detected. This signal is due to
molecules that are 
produced by the trapping light. It is subtracted as background 
from each pump-probe spectrum.  

In Fig.~\ref{fig:Kantenscan1} the 'detuning' of the spectral cut
position with respect to the atomic resonance increases from top to
bottom. A decreasing mean ion signal at positive time 
delays is observed for increasing cutoff position  for
excitation close to both the D1 and D2 line resonance. 
Moreover, the characteristic oscillations at positive time delay
decrease in their period and amplitude as the cutoff position is moved
further away from the resonance.
The decrease in oscillation period is explained in terms of the
instantaneous frequency of the pump pulse. 
A linear dependence of the oscillation frequency on the frequency of
the cutoff position was recorded, cf.  Fig.~2 of
Ref.~\cite{SalzmannPRL08}. This is due to the laser field
oscillating with an instantaneous frequency that corresponds to the
cutoff position, cf. Sec.~\ref{sec:cohtrans}.
The decrease in amplitude with increasing cutoff detuning 
is attributed to the most important
frequencies being excluded from the pulse. Moreover, 
the pulse energy is reduced 
when more frequency components are removed from 
the pump pulse spectrum. Then the instantaneous intensity and hence the
coupling between the field and  
and the molecular dipole become smaller. 

One might expect nuclear dynamics of the excited state wavepacket
to show up as oscillations in the molecular ion signal. However, the
vibrational periods of the excited state levels that are populated,
cf. Fig.~\ref{fig:vibstate}, are at least one order of magnitude larger than the
oscillation periods observed in Fig.~\ref{fig:Kantenscan1}. Moreover,
one would expect to see wave packet revivals in the vibrational
dynamics which have not been observed even when scanning the delay for
very long times (up to 250$\,$ps). In the calculations,
vibrational dynamics should yield slightly different
oscillation periods for the various potentials due to the different effective
$C_3$ coefficients. However, identical oscillation periods were
obtained for all potentials, independent of their character. All these
observations clearly rule out molecular vibrational dynamics as origin
of the observed oscillations and support their purely electronic
character.

\begin{figure*}[tb]
\includegraphics[width=17cm]{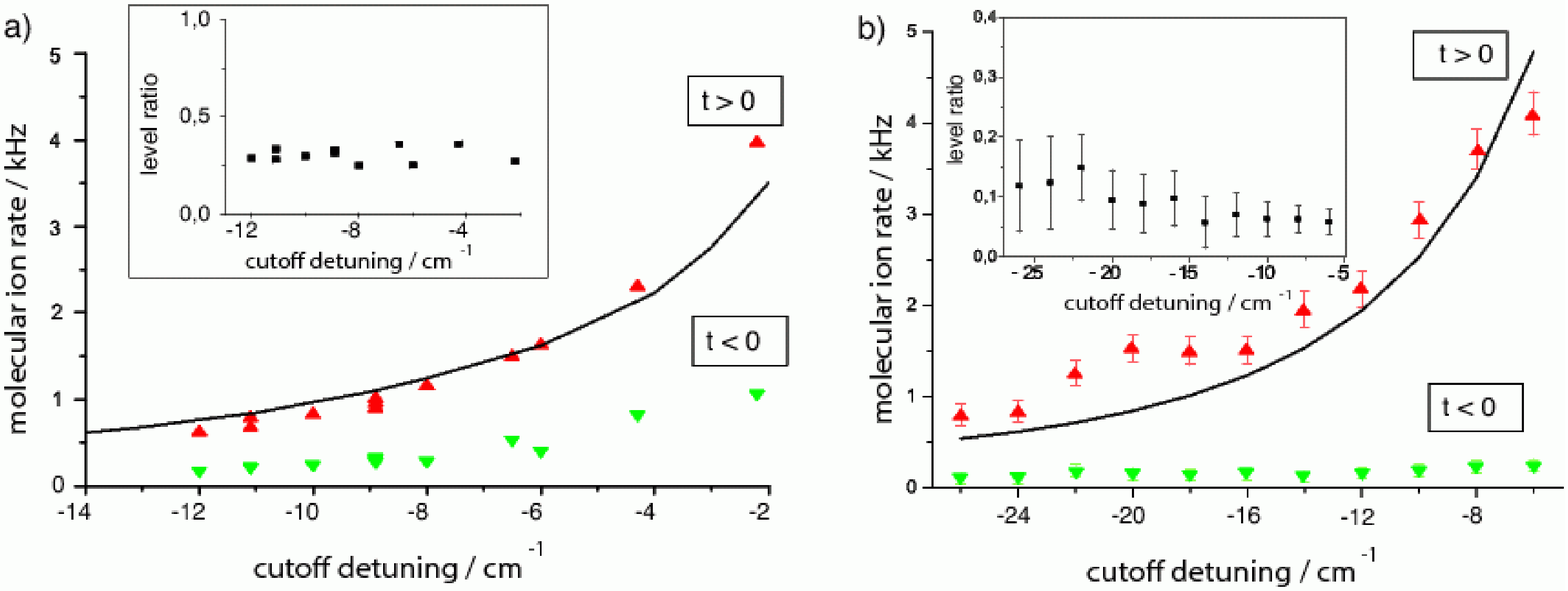}
\caption{\label{fig:fig9a_levels_vs_cut}
  (color online)
  Averaged asymptotic Rb$_{2}^{+}$ rates at positive ($t>0$) and
  negative ($t<0$) delay versus spectral cutoff position. The
  experimental results 
  are in good agreement with theory for both D1 (a)
  and D2 (b) resonance. The insets show the ratio of the asymptotic
  signals at positive and negative delay for a given cutoff position. 
} 
\end{figure*}
Fig.~\ref{fig:fig9a_levels_vs_cut} displays the 
variation of the averaged asymptotic pump-probe signals with the
cutoff position. Asymptotic refers to delay times for positive
delay ($t>0$) where the oscillatory behavior is fully damped and
to delay times before the signal rise  for 
negative delay ($t<0$).
The measured data are shown as upward triangles for positive delay and as
downward triangles for negative delay. The results of the calculation
were scaled to match the experimental data and 
are represented by a grey line (positive delay only). 
The insets display the ratio of the
asymptotic pump-probe signals at positive and negative delay versus
the cutoff position. This
ratio is obtained to be about  0.3 at the D1 line and about 0.1 at the
D2 line. As discussed in detail in Ref.~\cite{part1}, the  nearly
constant ratio indicates that the signal at negative delay, when the
probe precedes the pump pulse is related to molecules that were
photoassociated by a previous pump pulse and have undergone 
spontaneous decay to the electronic ground state.
The experimental result for positive delay is well reproduced by the
calculation. In particular, in accordance with the results of
Fig.~\ref{fig:Kantenscan1} 
an increase of the excited state population is observed as
the cutoff position approaches the resonance frequency. 

\begin{figure*}[tb]
\includegraphics[width=17cm]{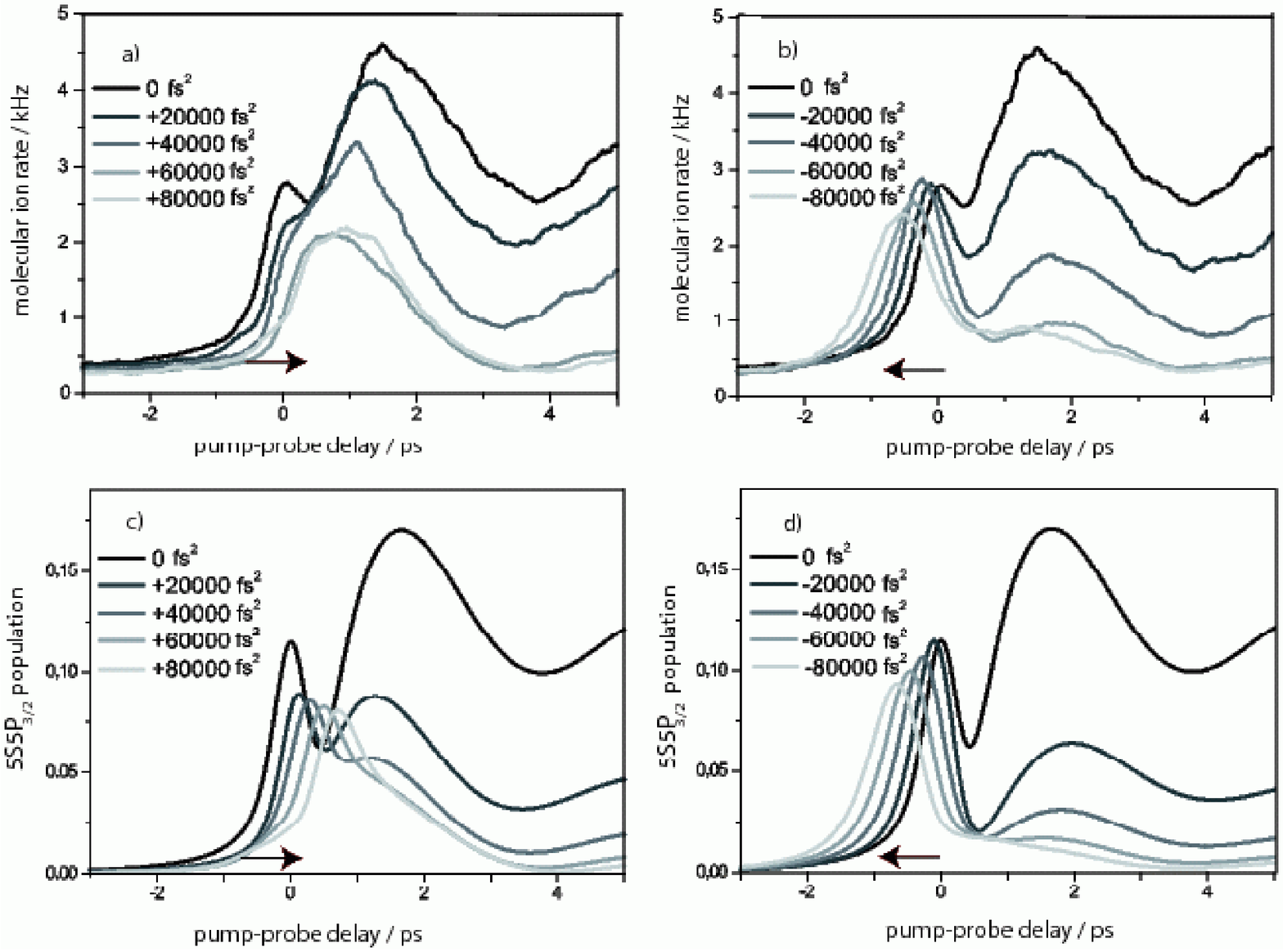}
\caption{\label{fig:pump-probe-chirp}
  (color online)
  Pump-probe spectra for different quadratic spectral phase
  shifts, i.e. linear chirps and a pump pulse cutoff position
  of 8$\,$cm$^{-1}$ below the D2 line resonance
  frequency. Experimental (theoretical) results are shown in top
  (bottom) panels.   As indicated by the arrows, a positive chirp (a,
  c) induces a shift of the zero delay peak 
  to positive time delays, whereas a negative chirp (b, d) 
  causes the opposite effect.}
\end{figure*}

The influence of a linear frequency chirp on the pump-probe signals is
discussed with the help of Fig.~\ref{fig:pump-probe-chirp}.
Measured (calculated) pump-probe spectra are shown for different
positive chirps in Fig.~\ref{fig:pump-probe-chirp}a (c) and for 
negative chirps in Fig.~\ref{fig:pump-probe-chirp}b (d) for an
excitation near the D2 line resonance and a cutoff position of
8$\,$cm$^{-1}$ below the resonance frequency.
The spectral cutoff in combination with the applied quadratic phase
shift causes strong distortions 
of amplitude and frequency behavior during the pulse compared to the
transform-limited and spectrally cut, but unchirped pulses. A
frequency chirp conserves the spectral bandwidth but prolongs the
pulse. Since the pulse energy remains unchanged, 
the peak intensities are decreased. This leads to
a reduction of the detected ion signal compared to an unchirped
pump pulse for both negative and positive time delays, cf. the black
and grey lines in Fig.~\ref{fig:pump-probe-chirp}.
Coherent transients are present also in the 
pump-probe spectra for chirped pump pulses. The chirp does 
not change their period, but it induces a shift of the zero delay peak,
as indicated by the arrows in 
Fig.~\ref{fig:pump-probe-chirp}. This behavior is well reproduced by
quantum dynamical calculations. For a positive (negative) chirp, the
larger instantaneous frequencies occur after (before) the smaller ones
during the pulse. The larger instantaneous frequencies are more likely
to cause the off-resonant excitation of population into the excited
state than the smaller ones. This explains the shift of the zero delay
peak toward later (earlier) times for positive (negative) chirps.

In previous calculations with chirped pulses, 
e.g. \cite{vala2001,ElianePRA04} an enhanced
efficiency was proposed for chirped pulses. Here, 
an enhancement of the excited state population by using linearly chirped
pump pulses was not observed.  This is due 
to the off-resonant character of the excitation,
cf. Sec.~\ref{sec:PA}. References~\cite{vala2001,ElianePRA04} describe
resonant excitation, i.e. only levels with free-bound
transition frequencies within the spectrum of the pulse were
addressed. Moreover, picosecond pulses were considered where some
vibrational dynamics occurs during the pulse. The proposed enhancement
then results from an interplay between the wavepacket rolling down the
potential and the coupling with the field. This cannot be expected in our case. 
Note that such an interplay between nuclear and electronic dynamics
has been recently observed for nanosecond pulses, demonstrating 
coherent control of ultracold collisions with 
frequency-chirped light \cite{WrightPRA07}.
In addition to reducing excitation efficiency, the linear chirps
affect the dynamics of the induced dipole. This can
be seen in the shift of the first peak relative to the oscillations 
indicated by the arrows in Fig.~\ref{fig:pump-probe-chirp} for
positive and negative chirps. 
As discussed above, the oscillations arise from the energy exchange
between the induced dipole and the electric field of the pulse. They
can  therefore be expected to be very sensitive to phase manipulations
of the pulse which also affect the relative phase between dipole and
field. 

\begin{figure*}[tb]
\begin{center}
\includegraphics[width=2 \columnwidth]{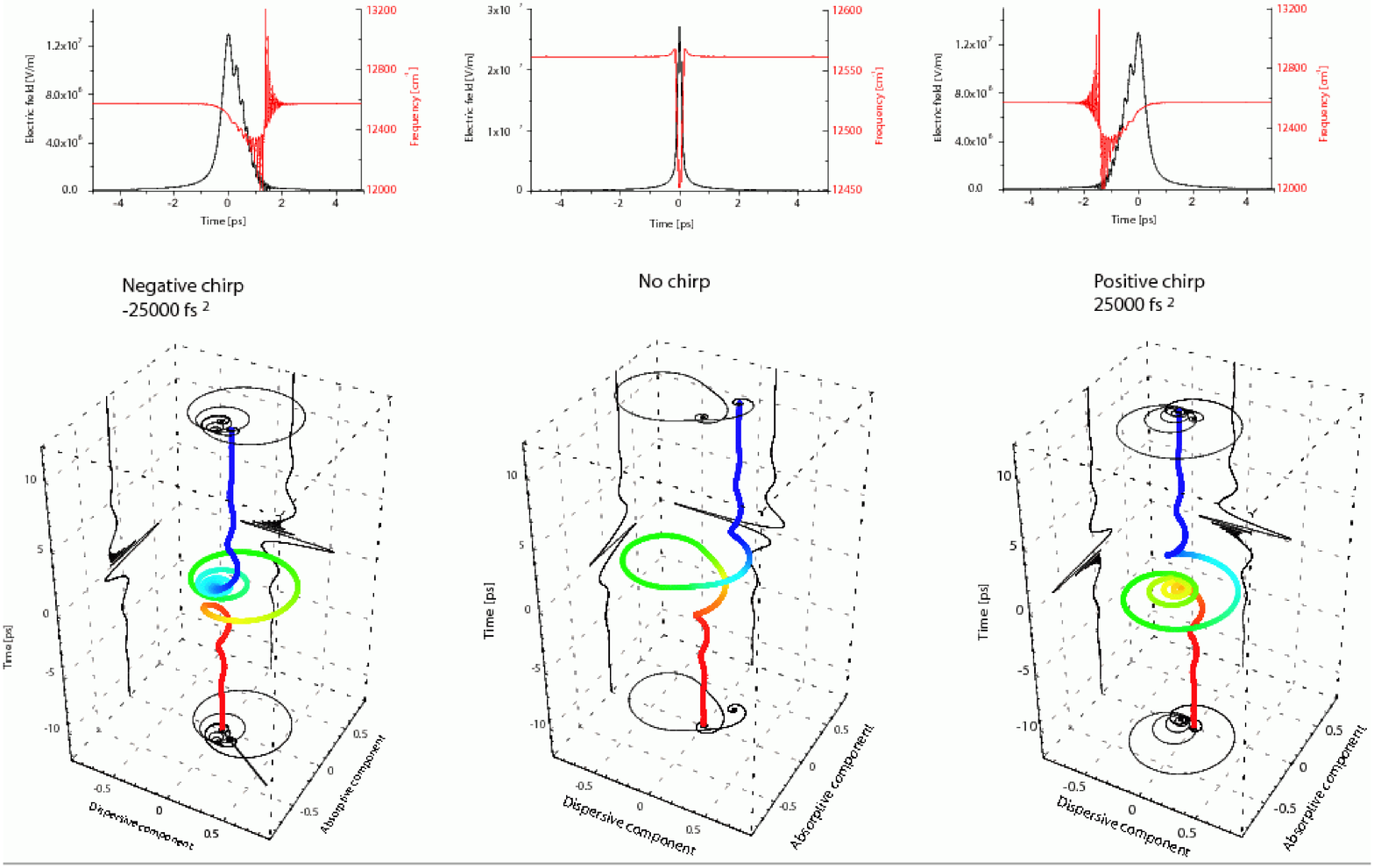}
\caption {
  (color online)
  Top panel: 
  Electric field (black lines) and instantaneous frequency (red line)
  of pulses with negative, no, and positive chirp.
  Bottom panel:
  Absorptive and dispersive components of the system Bloch vector as
  chirped, spectrally cut pump pulses interact with the system. Linear
  chirps result in different final phases of induced dipole and pulse
  field, causing phase shifts in the coherent transient modulations.}
\label{fig13_blochvectors.eps}
\end{center}
\end{figure*}
Since the molecular dynamics is reduced to the electronic dynamics,
i.e. to that of a two-level system, 
the interaction between the molecule and the field can also be
analyzed in terms of the evolution of the Bloch vector \cite{albert2008}. 
Fig.~\ref{fig13_blochvectors.eps} shows the  dispersive and absorptive 
components, $u(t)=\rho_{ge}(t)+\rho_{eg}(t)$ and
$v(t)=i(\rho_{ge}(t)-\rho_{eg}(t))$, of the Bloch vectors for chirped and 
unchirped pulses versus time 
\footnote{The inversion component, $\sigma_z(t)$, 
plotted vertically in the Bloch sphere, is small throughout the
interaction.}. For unchirped pulses, cf. middle graph in
Fig.~\ref{fig13_blochvectors.eps}, the Bloch vector moves merely
through a 
single loop in the $u-v$ plane during the main pulse peak. For chirped
pulses the Bloch vectors perform rapid rotations in the $u-v$ plane 
during the main pulse peak as the instantaneous frequency of the pulse
varies, cf. Fig.~\ref{fig10_dipole_interaction}a.
For negative (positive) chirps, cf. the left (right) graph of
Fig.~\ref{fig13_blochvectors.eps}, 
the rapid rotations occur before (after) $t=0$.
As a result the final position of the Bloch vector in the
$u-v$ plane is different for different sign of the chirp. 
The modulations on the absorptive component that are responsible for
the energy exchange between molecule and field then oscillate with a
phase offset compared to zero chirp.

\section{Conclusions}

\label{sec:concl}

We have presented theoretical calculations describing femtosecond
pump-probe photoassociation experiments
\cite{SalzmannPRL08,part1}. Two colliding rubidium atoms in their
electronic ground state are excited by the pump pulse into bound
levels and continuum states near the $5s+5p_{1/2}$ and $5s+5p_{3/2}$
asymptotes. They are ionized by a second, time-delayed probe pulse and
detected by a mass spectrometer. The model considers only the pump pulse
explicitly while accounting for the finite width of the
probe pulse by convolution with a Gaussian. The pump pulse was modeled
according to the experimental conditions. In particular, spectral
cuts of amplitude and quadratic spectral phases were considered. 
Photoassociation is observed as excitation of the two atoms into 
weakly bound molecular levels, mostly very close to the dissociation
limit. It is attributed mainly to off-resonant excitation. 
The spectral cutoff produces pulses in time with very long
tails giving rise to transient oscillations of the ion
signal, respectively excited state population. Coherent transients
were analyzed in terms of the field-induced dipole and the relative
phase between system and field. Pump pulses obtained for different cut
positions and linear chirps were employed. The oscillation period of
the coherent transients is directly related to the cutoff position.
A negative (positive) chirp simply shifts the
zero peak delay to shorter (larger) delay times.
Calculations were performed for a number of different excited state
potential energy curves correlating to both the $5s+5p_{1/2}$ and
$5s+5p_{3/2}$ dissociation limits yielding, however, identical oscillation
periods of the coherent transients. This indicates a purely electronic
character of the observed oscillations. The analysis of the two-level
dynamics  was further confirmed by inspection of the system's Bloch
vector. 

The experiments modelled in this paper have provided the first
evidence for femtosecond photoassociation. The off-resonant
excitation of molecular levels and the coherent transients both
clearly demonstrate  that the observed phenomena are beyond the sudden
approximation. However, the molecules created by the femtosecond pump
pulse are extremely weakly bound -- they do not provide a suitable
starting point for the application of a dump pulse in order to create
ultracold molecules in their electronic ground state. In future
experiments, the pulses should therefore be better
adapted to the timescales of the problem. In particular, a bandwidth
of a few wavenumbers should be best to create a spatially
localized excited state wavepacket
\cite{ElianePRA04} that can subsequently be dumped to the electronic
ground state \cite{KochPRA06a,KochPRA06b}. 

Theoretical estimates
predict a moderate photoassociation efficiency of about one molecule
per pulse for a magneto-optical trap (MOT) \cite{KochJPhysB06}. The efficiency is
limited by the comparatively small pair density at the distances where 
photoassociation is effective. Future experiments on short-pulse
photoassociation should therefore be combined with efforts to increase
the pair density. Two routes can be pursued towards this goal. (i) The
atoms can be cooled further to reach the quantum degenerate
regime. Compared to the highest possible densities in a MOT, about
three orders of magnitude can be gained this way. (ii) External fields
can be employed to enhance the pair density at short and intermediate
distances. For example, an increase in the photoassociation efficiency
of up to five orders of magnitude is predicted for magnetic field
control in Feshbach-optimized photoassociation
\cite{PellegriniPRL08}. 
Pursuing one of these routes or a combination thereof, an
efficiency of $10^3$ molecules per pulse or more is
to be expected in photoassociation with picosecond pulses.

\begin{acknowledgments}
This work was supported by the Deutsche Forschungsgemeinschaft in the
framework of SFB 450, SPP 1116, and the Emmy Noether programme
(RA and CPK).
\end{acknowledgments}


\end{document}